\documentclass[technologies,article,accept,moreauthors,pdftex,10pt,a4paper]{mdpi} 

\firstpage{1} 
\articlenumber{x}
\doinum{10.3390/------}
\pubvolume{xx}
\pubyear{2016}
\copyrightyear{2016}
\externaleditor{Academic Editor: name}
\history{Received: date; Accepted: date; Published: date}
\usepackage{amsfonts}
\usepackage{amssymb}
\usepackage{graphicx}
\usepackage{amsmath}
\usepackage{subfig}
\usepackage{epsfig}
\usepackage{bm}

 \theoremstyle{mdpi}
 \newcounter{thm}
 \newcounter{ex}
 \newcounter{re}
\Title{Correlation Plenoptic Imaging With \mbox{Entangled Photons}}

\Author{Francesco V. Pepe $^{1,2}$, Francesco Di Lena $^{2,3}$, Augusto Garuccio $^{2,3}$, Giuliano Scarcelli $^{4}$ and \mbox{Milena D'Angelo} $^{2,3,}$*}
\AuthorNames{F.V. Pepe, G. Scarcelli, A. Garuccio and M. D'Angelo}

\address{%
$^{1}$ \quad Museo Storico della Fisica e Centro Studi e Ricerche ``Enrico Fermi'', I-00184 Roma, Italy; francesco.pepe@ba.infn.it\\
$^{2}$ \quad Istituto Nazionale di Fisica Nucleare (INFN), Sezione di Bari, I-70126 Bari, Italy; francesco.dilena@uniba.it~(F.D.L.); augusto.garuccio@uniba.it (A.G.) \\
$^{3}$ \quad Dipartimento Interateneo di Fisica, Universit\`a degli studi di Bari, I-70126 Bari, Italy\\
$^{4}$ \quad Fischell Department of Bioengineering, University of Maryland, College Park, MD 20742, USA; scarc@umd.edu\\}

\corres{Correspondence: milena.dangelo@uniba.it; Tel.: +39-080-554-3217}

\abstract{Plenoptic imaging is a novel optical technique for three-dimensional imaging in a single shot. It is enabled by the simultaneous measurement of both the location and the propagation direction of light in a given scene. In the standard approach, the maximum spatial and angular resolutions are inversely proportional, and so are the resolution and the maximum achievable depth of focus of the 3D image. We have recently proposed a method to overcome such fundamental limits by combining plenoptic imaging with an intriguing correlation remote-imaging technique: ghost imaging. Here, we theoretically demonstrate that correlation plenoptic imaging can be effectively achieved by exploiting the position-momentum entanglement characterizing spontaneous parametric down-conversion (SPDC)  photon pairs. As a proof-of-principle demonstration, we shall show that correlation plenoptic imaging with entangled photons may enable the refocusing of an out-of-focus image at the same depth of focus of a standard plenoptic device, but without sacrificing diffraction-limited image resolution.}


\begin{document}

\section{Introduction}

Plenoptic imaging, also known as light-field or integral imaging, is a novel optical imaging modality \cite{adelson}. Its key principle is to record the three-dimensional light field of a given scene by measuring both the location and the propagation direction of the incoming light. In particular, several images of the scene, one for each propagation direction of light within the scene, are acquired in a single shot. On one hand, such images correspond to the required viewpoints enabling the three-dimensional reconstruction of the scene. In fact, plenoptic imaging is the simplest method of 3D imaging with the present technological means \cite{3dimaging,microscopy2,microscopy4}. On the other hand, the available angular information also enables the simplification of low-light shooting: The acquired images can be combined, in post-processing, to give an overall image characterized by the same depth of field of the $N$ original images, but a signal-to-noise ratio $N$ times larger \cite{ng}. 
\newpage
Plenoptic imaging is currently used in digital cameras enhanced by refocusing capabilities  \cite{website1,website2,website3}; in fact, in photography, plenoptic imaging highly simplifies both auto-focus and low-light \mbox{shooting \cite{ng}.} A plethora of innovative applications in 3D imaging and sensing \cite{3dimaging,waller_turb}, \mbox{stereoscopy~\cite{adelson,muenzel,levoy}} and microscopy \cite{microscopy1,microscopy2,microscopy3} are also being developed. In particular, high-speed large-scale 3D functional imaging of neuronal activity has been demonstrated \cite{microscopy4}. 

However, the potentials of plenoptic imaging are strongly limited by the inherent inverse proportionality between image resolution and maximum achievable depth of field. In fact, plenoptic imaging has so far been implemented by inserting a microlens array in the native image plane, while moving the sensor array behind the microlenses. The image of the scene is  reproduced on the microlenses, which thus define the spatial resolution of the acquired image. Each microlens also serves for reproducing, on the sensor array, an image of the camera lens, thus providing the angular information associated with each imaging pixel \cite{ng}. As a result, a trade-off between spatial and angular resolution is built in the plenoptic imaging process. To recover the lost resolution, signal processing and deconvolution have been implemented  \cite{microscopy2,microscopy4,waller,spatioangular,imageformation,superres}.

We have recently proposed a novel approach to plenoptic imaging, named correlation plenoptic imaging (CPI), which exploits the spatio-temporal second-order correlation typical of chaotic light sources to beat the strong coupling between spatial and angular resolution, as imposed to standard plenoptic imaging devices \cite{plenoptic_prl,plenoptic_qmqm}. From a fundamental standpoint, the plenoptic application has been the first physical context where the counterintuitive properties of chaotic light (namely, the coexistence of momentum and position correlation \cite{ferri}) are effectively used to beat intrinsic limits of standard imaging systems. From a practical standpoint, our protocol has been shown to dramatically enhance the potentials of plenoptic imaging. However, in contrast with chaotic light \cite{torino}, correlation imaging based on entangled photons has been shown to enable sub-shot-noise imaging \cite{torino_nature}, as required by biomedical and security applications. Hence, in this paper, we investigate the possibility of performing CPI with entangled photons, or twin beams, from spontaneous parametric down-conversion (SPDC) \cite{klyshko}. We show that the peculiar momentum-momentum and position-position correlations typical of such EPR entangled systems \cite{ghost2,laserphys} can be simultaneously exploited to substantially weaken the connection between spatial resolution and depth of field typical of standard plenoptic imaging. 

\begin{figure}[H]
\centering
{\includegraphics[width=0.9\textwidth]{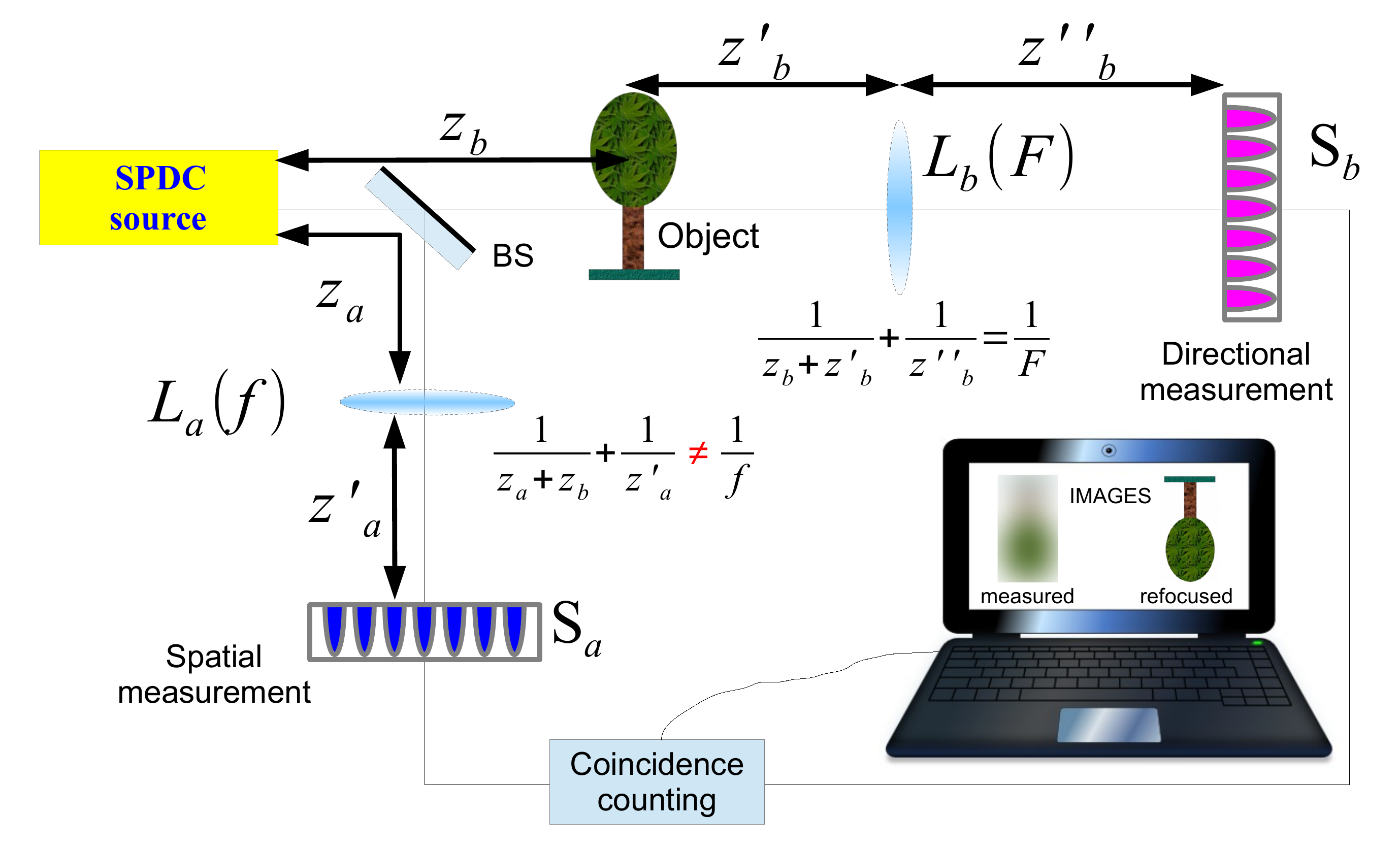}}
\caption{Schematic setup for correlation plenoptic imaging with entangled photons from SPDC. Signal and idler beams emitted from the SPDC source impinge on a beam-splitter (BS). Both beams are split into a reflected path $a$ and a transmitted path $b$. The reflected beam propagates toward the lens $L_a$ of focal length $f$ and is refracted toward the high resolution sensor array $\mathrm{S}_a$. The transmitted beam propagates through the object, playing the role of the desired scene, and is collected by the lens $L_b$ of focal length $F$ before being detected by the high-resolution sensor array $\mathrm{S}_b$. The two sensors are connected to a coincidence counting circuit. On one hand, distances $z_b, z'_b, z''_b$ are chosen in such a way that the source and the sensor $\mathrm{S}_b$ are in conjugate planes of the lens $L_b$.
On the other hand, distances $z_a$ and $z'_a$ are such that, when the two-photon thin-lens equation $1/(z_b+z_a)+1/z'_a=1/f$ is satisfied, a ghost image of the object is retrieved on sensor $\mathrm{S}_a$, triggered by sensor $\mathrm{S}_b$.}\label{fig:setup}
\end{figure}

The proposed setup for CPI with entangled photons from SPDC is reported in Figure \ref{fig:setup}. In view of plenoptic imaging, the setup must enable the parallel acquisition of several images of the given scene, one for each propagation direction of light. In fact, as we shall soon demonstrate, the sensor array $\mathrm{S}_a$ retrieves $N$ \textit{coherent ghost images} of the object by means of correlation measurement with each of the pixels of the sensor array $\mathrm{S}_b$. Such images represent different viewpoints of the desired scene. This is quite intuitive considering sensor $\mathrm{S}_b$ reproduces the image of the light source. Hence, each coherent ghost image is associated with a different illumination of the object. Interestingly, the single lens $L_b$ replaces the microlens array required in standard plenoptic imaging. In summary, the basic idea of CPI is to replace with a single lens and two separate sensors, the complex system composed of the microlens array followed by a single sensor; spatial and angular measurements are thus physically decoupled, enabling a significant weakening of the inverse proportionality between spatial and angular resolution characterizing standard plenoptic imaging devices.

\section{Theoretical Analysis}

\subsection{Background}\label{sect:mat}

The coincidence detection of entangled photons from SPDC is described by the second order Glauber correlation function \cite{scully}:
\begin{equation}\label{G2}
G^{(2)}(\bm{r}_a,\bm{r}_b;t_a,t_b)\! = \langle \Psi | E^{(-)}_a(\bm{r}_a,t_a) E^{(-)}_b (\bm{r}_b,t_b) E^{(+)}_b(\bm{r}_b,t_b) E^{(+)}_a(\bm{r}_a,t_a) | \Psi \rangle,
\end{equation}
where 
\begin{equation}\label{field}
E^{(+)}_j (\bm{r}_j,t_j) = \int d\omega \int d\bm{\kappa}\, a_{\bm{k}} e^{-i\omega t_j} g_j(\bm{r}_j,\bm{k}),
\end{equation}
is the positive-energy part of the electric field at sensor $j$ (with $j=a,b$), placed in $\bm{r}_j=(\bm{\rho}_j,z_j)$, $t_j$ the time of the detection, $\omega$ is the frequency and $\bm{k}=(\bm{\kappa},\omega/c)$ the wave vector of the detected radiation, $g_j$ is the Green's function propagating the field mode $\bm{k}$ from the source to the sensor. The negative-energy part $E^{(-)}_j (\bm{r}_j,t_j)$ of the electric field is the Hermitian conjugate of the field $E^{(+)}$ of \mbox{Equation (\ref{field}).} A scalar approximation for the electric field has been assumed, which physically corresponds to considering a fixed polarization of light. The positive and negative-energy parts of the electric field involve the photon annihilation and creation operators ($a_{\bm{k}}$ and $a_{\bm{k}}^{\dagger}$), respectively, associated with wave vector $\bm{k}$. The expectation value in Equation (\ref{G2}) is taken over the two-photon signal-idler state produced by SPDC \cite{spdc1,spdc2,spdc3}:
\begin{equation}
| \Psi \rangle = \mathcal{N} \int d\nu s(L D \nu) \int d\bm{\kappa}_i d\bm{\kappa}_s h_{\mathrm{tr}} (\bm{\kappa}_i + \bm{\kappa}_s) a^{\dagger}_{\bm{k}_i} a^{\dagger}_{\bm{k}_s} |0\rangle,
\end{equation}
where $\mathcal{N}$ is a normalization constant, $\nu$ is the detuning with respect to the central frequency of signal and idler $\Omega_s=\Omega_i=\omega_p/2$ , which is linked by phase matching to the central frequency of the pump laser $\omega_p$, $L$ is the length of the SPDC crystal, $D$ is the difference between the inverse group velocities of signal and idler, $s(L D \nu)$ is the spectrum of the SPDC biphoton \cite{SPDC_spectrum1,SPDC_spectrum2}, and $h_{\mathrm{tr}}$ is the Fourier transform of the pump transverse profile:
\begin{equation}\label{f}
\mathcal{F}(\bm{\rho}) = \int d\bm{\kappa} e^{i\bm{\kappa}\cdot\bm{\rho}} h_{\mathrm{tr}} (\bm{\kappa}).
\end{equation}

We have assumed, for simplicity, degenerate SPDC radiation, but the result can be easily generalized to the non-degenerate situation \cite{GI-2colour_th,GI-2colour_exp}. Without loss of generality, we shall further assume the source to be monochromatic, in such a way that the time dependence of the correlation function will not be relevant. By employing the canonical commutation relations $[a_{\bm{k}},a_{\bm{k}'}]=0$ and $[a_{\bm{k}},a_{\bm{k}'}^{\dagger}]=\delta(\bm{k}-\bm{k}')$, with $\delta$ the Dirac delta distribution, and the inversion symmetry of the Fourier transform of the transverse pump profile $h_{\mathrm{tr}}(\bm{\kappa})=h_{\mathrm{tr}}(-\bm{\kappa})$, the spatial part of the two-photon correlation function reads:
\begin{equation}\label{G2_g}
\Gamma (\bm{\rho}_a,\bm{\rho}_b) = \left| \int d\bm{\kappa}_a \int d\bm{\kappa}_b g_a (\bm{\rho}_a,\bm{\kappa}_a) g_b (\bm{\rho}_b,\bm{\kappa}_b) h_{\mathrm{tr}} (\bm{\kappa}_a+\bm{\kappa}_b) \right|^2,
\end{equation}
up to irrelevant constants. This result indicates the strong coupling between the two remote sensors, as enabled by the momentum-momentum entanglement characterizing SPDC biphotons.

Let us now evaluate the propagators in the two arms of the setup depicted in Figure \ref{fig:setup}; we shall assume for simplicity the lenses to be diffraction-limited. In arm $a$, light propagates in free space for a distance $z_a$ from the source to the lens $L_a$ and is then detected by the sensor $\mathrm{S}_a$, placed at \mbox{a distance $z_a'$} from the lens. In the paraxial approximation, propagation of a field with frequency $\Omega\simeq c k_z$ in free space from $(\bm{\rho}_1,z_1)$ to $(\bm{\rho}_2,z_2)$ is described by the function \cite{goodman}:
\begin{equation}
\mathcal{G}(\bm{\rho}_2 - \bm{\rho}_1, z_2 - z_1) = \frac{-i \Omega e^{i \frac{\Omega}{c} (z_2-z_1)}}{2\pi c (z_2-z_1)} G(\bm{\rho}_2 - \bm{\rho}_1 )_{\left[\frac{\Omega}{c (z_2-z_1)}\right]} 
\end{equation}
with $G(\bm{x})_{[y]} = e^{ i y |\bm{x}|^2/2}$. Knowing the initial field $E(\bm{\rho}_1)$, one can determine the final field $E(\bm{\rho}_2) = \int d\bm{\rho}_1 E(\bm{\rho}_1) \mathcal{G}(\bm{\rho}_2 - \bm{\rho}_1, z_2 - z_1)$. Propagation through a lens of focal length $f$ is described by $G(\bm{\rho}_l)_{[-\Omega/(cf)]}$. Hence, the propagator associated with arm $a$ of the setup reads:
\begin{equation}
\begin{aligned}\label{ga}
g_a (\bm{\rho}_a,\bm{\kappa}_a) & = \mathcal{C}_a(z_a,z_a') \int d\bm{\rho}_s \int d\bm{\rho}_{\ell} e^{i\bm{\kappa}_a\cdot\bm{\rho}_s} 
G(\bm{\rho}_{\ell}-\bm{\rho}_s)_{\left[ \frac{\Omega}{cz_a} \right]} G(\bm{\rho}_{\ell})_{\left[ -\frac{\Omega}{cf} \right]} G(\bm{\rho}_a-\bm{\rho}_{\ell})_{\left[ \frac{\Omega}{cz_a'} \right]}  \\
& = \mathcal{C'}_a(z_a,z_a') G(\bm{\rho}_a)_{\left[ \frac{\Omega}{c} \left( \frac{1}{z_a} - \frac{\zeta(z_a,z_a')}{{z_a'}^2} \right) \right]} 
\int d\bm{\rho}_s  e^{i\bm{\kappa}_a\cdot\bm{\rho}_s} G(\bm{\rho}_s)_{\left[ \frac{\Omega}{c z_a} \left( 1 - \frac{\zeta(z_a,z_a')}{z_a} \right) \right]} e^{- \frac{i\Omega \zeta(z_a,z_a')}{c z_a z_a'} \bm{\rho}_s \cdot \bm{\rho}_a },
\end{aligned}
\end{equation}
where
\begin{equation}\label{zeta}
\zeta(z_a,z_a') = \left( \frac{1}{z_a}+\frac{1}{z_a'}-\frac{1}{f} \right)^{-1},
\end{equation}
$\bm{\rho}_s$ and $\bm{\rho}_{\ell}$ are transverse coordinate on the source and the lens $L_a$ plane, respectively, and $\mathcal{C}_a, \mathcal{C'}_a$ contain irrelevant constants. In arm $b$, light propagates for a distance $z_b$ from the source to the object which represents the desired scene to image, then for a distance $z'_b$ from the object to lens $L_b$, and a further distance $z''_b$ before being detected by the sensor $\mathrm{S}_b$. By indicating with $A$ the aperture function of the object, and assuming the focusing condition $1/(z_b'+z_b'')+1/z_b = 1/F$ to be satisfied, the propagator associated with arm $b$ of the setup reads:
\begin{equation}
{\small\begin{aligned}\label{gb}
g_b (\bm{\rho}_b,\bm{\kappa}_b) = & \,\mathcal{C}_b(z_b,z_b') \int d\bm{\rho}_s \int d\bm{\rho}_o \int d\bm{\rho}'_{\ell} e^{i\bm{\kappa}_a\cdot\bm{\rho}_s} A(\bm{\rho}_o)
G(\bm{\rho}_o-\bm{\rho}_s)_{\left[ \frac{\Omega}{cz_b} \right]} G(\bm{\rho}'_{\ell}-\bm{\rho}_o)_{\left[ \frac{\Omega}{cz'_b} \right]}  \\ & \times G(\bm{\rho}_{\ell})_{\left[ -\frac{\Omega}{cF} \right]} G(\bm{\rho}_b-\bm{\rho}_{\ell})_{\left[ \frac{\Omega}{cz_b''} \right]} \\
= & \,\mathcal{C}'_b(z_b,z_b') G(\bm{\rho}_b)_{\left[ \frac{\Omega}{cz_b''} \left( 1 - \frac{1}{z_b''} \left( \frac{1}{z_b'}+\frac{1}{z_b''}-\frac{1}{F} \right)^{-1} \right) \right]} \int d\bm{\rho}_s d\bm{\rho}_o e^{i\bm{\kappa}_a\cdot\bm{\rho}_s} G(\bm{\rho}_s)_{\left[ \frac{\Omega}{cz_b} \right]} A(\bm{\rho}_o) e^{- \frac{i\Omega}{cz_b} \left(\bm{\rho}_s + \frac{\bm{\rho}_b}{M} \right)\cdot \bm{\rho}_o },
\end{aligned}}
\end{equation}
where $\bm{\rho}_o$ and $\bm{\rho'}_{\ell}$ are transverse coordinate on the object and the lens $L_b$ planes, respectively, \mbox{$M=z_b''/(z_b+z_b')$} is the magnification of the image of the source on the sensor array $\mathrm{S}_b$, and $\mathcal{C}_b, \mathcal{C'}_b$ contain irrelevant constants.

By inserting in Equation (\ref{G2_g}) the Green's function given by Equations (\ref{ga}) and (\ref{gb}), and the laser pump profile on the SPDC crystal, as defined in Equation (\ref{f}), one finds that the second order correlation function associated with signal-idler pairs from SPDC is given by the plenoptic \mbox{correlation function:} \begin{align}\label{Gamma}
\Gamma (\bm{\rho}_a,\bm{\rho}_b) = & \,\, \mathcal{K}(z_a,z_a',z_b,z_b') \Biggl| \int  d\bm{\rho}_o A(\bm{\rho}_o) 	\int d\bm{\rho}_s  
 \mathcal{F}(\bm{\rho}_s) G(\bm{\rho}_s)_{\left[ \frac{\Omega}{c} \left[ \frac{1}{z_b} + \frac{1}{z_a} \left(1 - \frac{\zeta(z_a,z_a')}{z_a} \right) \right] \right]} \nonumber \\ 
& \, \, e^{- \frac{i\Omega \zeta(z_a,z_a')}{c z_a z_a'} \bm{\rho}_s \cdot \bm{\rho}_a } e^{- \frac{i\Omega}{cz_b} \left(\bm{\rho}_s + \frac{\bm{\rho}_b}{M} \right)\cdot \bm{\rho}_o} 
\Biggr|^2,
\end{align}
where $\mathcal{K}$ contains irrelevant constants.

%%%%%%%%%%%%%%%%%%%%%%%%%%%%%%%%%%%%%%%%%%
\subsection{Plenoptic Properties of the Correlation Function and Refocusing Capability}

As shown in Equation (\ref{Gamma}), the proposed CPI protocol is theoretically described by a second order correlation function encoding both spatial and angular information, hence, characterized by the key re-focusing capability typical of plenoptic imaging.

To develop an intuition about the result of Equation (\ref{Gamma}), we consider the simple case in which the distance between the object and the source $z_b=z_{b F}$ satisfies the two-photon thin lens \mbox{equation \cite{laserphys,pittman}}:
\begin{equation}\label{thinlens_GI}
\frac{1}{z_a+z_{b F}} + \frac{1}{z_a'} = \frac{1}{f}.
\end{equation}

In this case, by integrating the result of Equation (\ref{Gamma}) over the whole sensor array $\mathrm{S}_b$, one gets the standard (\textit{incoherent}) ghost image of the object, magnified by a factor of $m=\frac{z_a'}{z_a+z_{b F}}$, namely \cite{laserphys,pittman},
\begin{equation}\label{Sigma_GI}
\Sigma_F(\bm{\rho}_a) = \int d{\bm{\rho}_b} \Gamma (\bm{\rho}_a,\bm{\rho}_b) \propto \int d{\bm{\rho}_o} |A(\bm{\rho}_o)|^2 \left|h_{\mathrm{tr}} \left[\frac{\Omega}{cz_{bF}} \left(\bm{\rho}_o + \frac{\bm{\rho}_a}{m} \right) \right] \right|^2, 
\end{equation}
where $h_{\mathrm{tr}}$ is the Fourier transform of the laser pump profile, as defined in Equation (\ref{f}). The above result is valid in the hypothesis that $h_{\mathrm{tr}}$ is similar to or narrower than the Fourier transform of the imaging lens $L_a$. In fact, such incoherent ghost image is formally equivalent to the incoherent image one would obtain in a standard imaging system characterized by a point-spread function $h_{\mathrm{tr}}$ given by the Fourier transform of the imaging lens aperture function.

However, the second order correlation function of Equation (\ref{Gamma}) can do much better than standard ghost imaging: The deep physical difference arises from the coherent nature of the ghost image it describes, 
\begin{equation}\label{Gamma_cGI}
\Gamma_F (\bm{\rho}_a,\bm{\rho}_b) =
\mathcal{K}(z_a,z_a',z_{b F},z_b') \left| \int  d\bm{\rho}_o A(\bm{\rho}_o)   h_{\mathrm{tr}} \left[\frac{\Omega}{c z_{b F}} \left(\bm{\rho}_o + \frac{\bm{\rho}_a}{m} \right) \right]      e^{- \frac{i\Omega}{c z_{b F}} \frac{\bm{\rho}_b}{M} \cdot \bm{\rho}_o }  \right|^2 ,
\end{equation}
that is obtained from the general expression (\ref{Gamma}), when the focusing condition in Equation ~(\ref{thinlens_GI}) holds.

\begin{figure}[H]
\centering
{\includegraphics[width=0.8\textwidth]{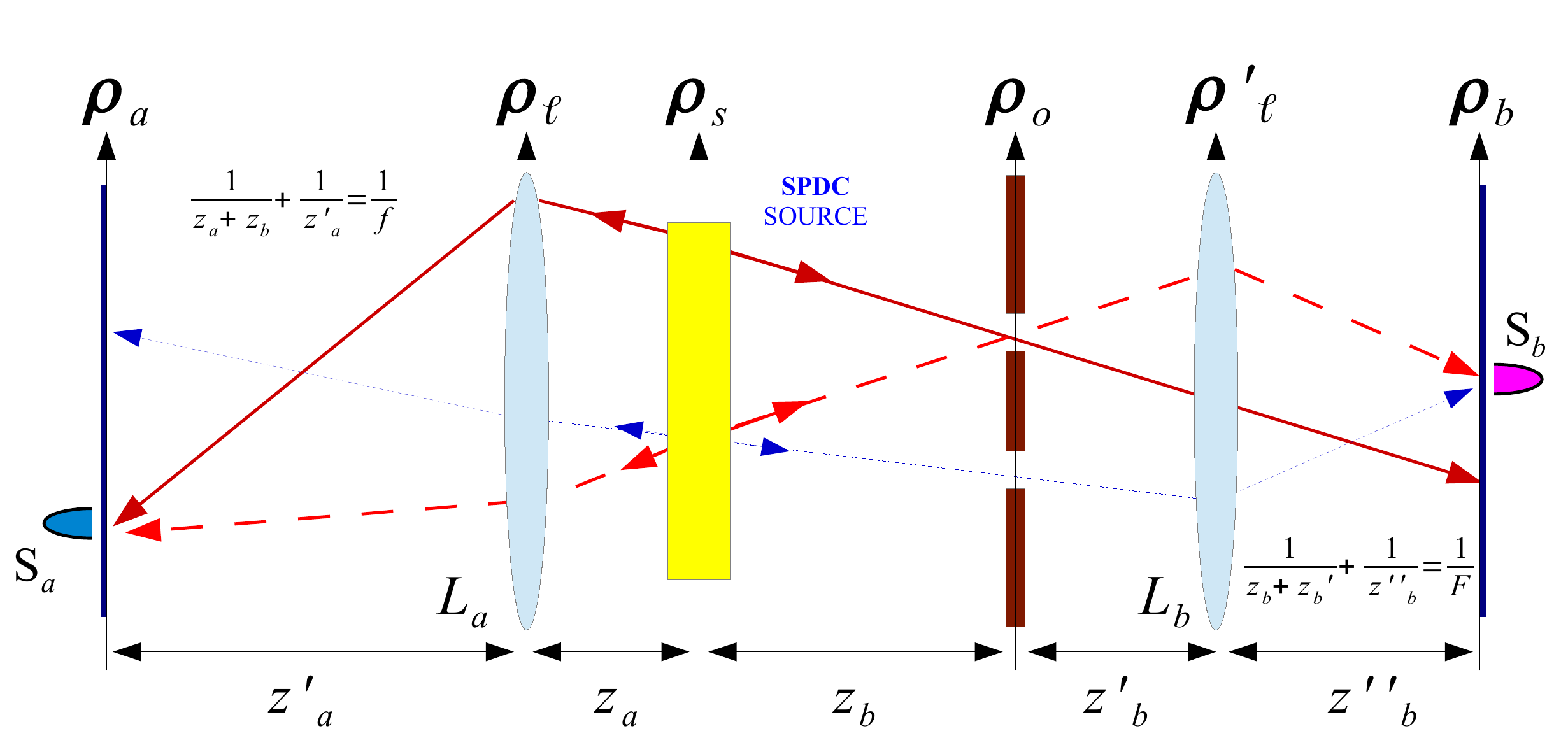}}
\caption{Unfolded version, or Klyshko picture, of the schematic setup reported in Figure~\ref{fig:setup}, in the case in which the ghost image of the object is focused on the sensor $\mathrm{S}_a$. On one hand, by means of coincidence detection, the lens $L_a$ reproduces on the sensor $\mathrm{S}_a$ the ghost image of the object. On the other hand, the source and the sensor $\mathrm{S}_b$ are in conjugate planes of the lens $L_b$. However, based on the advanced-wave model proposed by Klyshko, the effect can be understood by treating sensor $\mathrm{S}_b$ as the light source and the SPDC source as a simple mirror. The solid and the dashed lines represent two-photon amplitudes that pass through the same slit; hence, at second order, they are focused in the same point of sensor $\mathrm{S}_a$. The dashed and the dotted two-photon amplitudes are emitted by the same source point and are thus focused in the same point of sensor $\mathrm{S}_b$.}\label{fig:klyshko}
\end{figure}

The coherence of such ghost image is the immediate consequence of measuring coincidences between the \textit{spatial} sensor $\mathrm{S}_a$ and any single pixel of the \textit{angular} sensor $\mathrm{S}_b$. This can be better understood in terms of the Klyshko picture \cite{pittman} reported in Figure \ref{fig:klyshko}: The light illuminating the object and contributing to the coincidence detection between any two pairs of pixels $\bm{\rho}_a$ and $\bm{\rho}_b$ has a well defined propagation direction (\emph{i.e.}, it is spatially coherent). As made clear from Figure \ref{fig:klyshko}, the Klyshko picture also enables the interpretation of the proposed setup for CPI with entangled photons as a sort of \textit{correlation} pinhole camera. Such a perspective helps developing an intuition about the analogy between the proposed scheme and standard plenoptic imaging, as well as understanding the role played by the sensor $\mathrm{S}_b$ in retrieving the angular information about the two-photon light field. \mbox{In fact,} due to the quasi one-to-one correspondence between points on the sensor $\mathrm{S}_b$ and points on the source, one can trace, in post-processing, the geometrical ray connecting each point of the source with each point of the object. This leads to the peculiar refocusing and 3D imaging capabilities of \mbox{plenoptic imaging.}

Now, to explicitly demonstrate this last point and better highlight the plenoptic properties of the second-order correlation function of Equation (\ref{Gamma}), we shall consider the more general out-of-focus situation ($z_b \neq z_{bF} $) and rewrite it as a product of the pump profile $\mathcal{F}$ and the object aperture function $A$ with the phase factor $e^{i \frac{\Omega}{c} \varphi(\bm{\rho}_o,\bm{\rho}_s;\bm{\rho}_a,\bm{\rho}_b)}$, with:
\begin{equation}\label{phase}
\varphi(\bm{\rho}_o,\bm{\rho}_s;\bm{\rho}_a,\bm{\rho}_b)= \left[ \frac{1}{z_b} + \frac{1}{z_a} \left(1 - \frac{\zeta(z_a,z_a')}{z_a} \right) \right] \frac{|\bm{\rho}_s|^2}{2} - \frac{\zeta(z_a,z_a')}{z_a z_a'} \bm{\rho}_s \cdot \bm{\rho}_a - \frac{1}{z_b} \left(\bm{\rho}_s + \frac{\bm{\rho}_b}{M} \right)\cdot \bm{\rho}_o ,
\end{equation}
namely
\begin{equation}\label{Gamma_phi}
\Gamma (\bm{\rho}_a,\bm{\rho}_b) \propto \left| \int d\bm{\rho}_o A(\bm{\rho}_o) \int d\bm{\rho}_s \mathcal{F}(\bm{\rho}_s) e^{i\frac{\Omega}{c} \varphi(\bm{\rho}_o,\bm{\rho}_s;\bm{\rho}_a,\bm{\rho}_b)} \right|^2.
\end{equation}

The stationary points of the phase defined in Equation (\ref{phase}) enable us to determine the geometrical correspondence between points on the object and the source with points on the sensors $\mathrm{S}_a$ and $\mathrm{S}_b$, respectively. In particular, the stationarity of $\varphi$ with respect to $\bm{\rho}_s$ determines the object point that gives the predominant contribution to the integral of Equation (\ref{Gamma_phi}), that is:
\begin{equation}
\bm{\rho}_o = - \frac{z_b}{z_{bF}} \frac{\bm{\rho}_a}{m} - \frac{\bm{\rho}_b}{M} \left( 1 - \frac{z_b}{z_{bF}} \right),
\end{equation}
where the identity $\zeta(z_a,z_a')=(z_{bF}+z_a)z_a/z_{bF}$ has been used. When the focusing condition of Equation (\ref{thinlens_GI}) is satisfied, this object point becomes independent of the specific sensor pixel $\bm{\rho}_b$. Hence, the focused ghost image is not sensitive to the change of perspective enabled by the high resolution of the angular sensor $\mathrm{S}_b$. On the other hand, the stationarity of $\varphi$ with respect to $\bm{\rho}_o$ yields the focusing of the source on the sensor $\mathrm{S}_b$:
\begin{equation}
\bm{\rho}_s = - \frac{\bm{\rho}_b}{M}.
\end{equation}

Thus, in the geometrical optics limit, the second order correlation function of Equation (\ref{Gamma_phi}) reduces to the product of the tilted and rescaled geometrical image of the object and the source profile:
\begin{equation}\label{Gamma_out}
\Gamma_G (\bm{\rho}_a,\bm{\rho}_b) \sim  \left| A \left[  - \frac{z_b}{z_{bF}} \frac{\bm{\rho}_a}{m} - \frac{\bm{\rho}_b}{M} \left( 1 - \frac{z_b}{z_{bF}} \right) \right] \right|^2 \left| \mathcal{F} \left(-\frac{\bm{\rho}_b}{M} \right) \right|^2.
\end{equation}

Interestingly, by properly rescaling the variable $\bm{\rho}_a$, the object can be completely decoupled from the source; in fact, the rescaled second order correlation function
\begin{equation}\label{Gamma_riscaled}
\Gamma^{\mathrm{ref}}_G \left[ \frac{z_{bF}}{z_b} \bm{\rho}_a + \frac{\bm{\rho}_b}{M} m \left( 1 - \frac{z_{bF}}{z_b} \right), \bm{\rho}_b\right] \sim \left|  \mathcal{F} \left(-\frac{\bm{\rho}_b}{M} \right) \right|^2 \left| A\left( - \frac{\bm{\rho}_a}{m} \right) \right|^2 ,
\end{equation}
gives the perfect geometrical image of the desired scene. Such rescaling is formally identical to the one employed both in standard plenoptic imaging \cite{ng} and in correlation plenoptic imaging with chaotic light \cite{plenoptic_prl,plemoptic_qmqm}.

Similar to standard plenoptic imaging, the signal to noise ratio of the refocused image can be improved by integrating the result of Equation (\ref{Gamma_riscaled}) over the whole sensor array $\bm{\rho}_b$, thus employing light coming from the whole light source:
\begin{equation}\label{Sigma_riscaled}
\Sigma^{\mathrm{ref}}(\bm{\rho}_a) = \int d\bm{\rho}_b \Gamma^{\mathrm{ref}} \left[ \frac{z_{bF}}{z_b} \bm{\rho}_a + \frac{\bm{\rho}_b}{M} m \left( 1 - \frac{z_{bF}}{z_b} \right), \bm{\rho}_b\right] .
\end{equation}

This result represents the refocused incoherent ghost image of an object placed at a generic distance $z_b$ from the source, and is thus the central result of the present paper.

The possibility of reconstructing the light field and refocusing an out-of-focus image, as reported in Equation ~(\ref{Sigma_riscaled}), lies on the accuracy with which both object and source points are in a one-to-one correspondence with points on sensors $\mathrm{S}_a$ and $\mathrm{S}_b$, respectively. We have already demonstrated that the Fourier transform of the transverse pump profile determines the object point spread function (see Equation (\ref{Sigma_GI})), with a spot size $\Delta\rho_a \sim m c z_{bF} / (\Omega D_s)$, where $D_s$ is the diameter of the pump profile. On the other hand, it is easy to check that the source is imaged with a point spread function given by the Fourier transform of the object aperture function. From Equation~(\ref{Gamma}), one can infer that a point on the source corresponds to a spot of width $\Delta\rho_b{\sim}M c z_b / (\Omega d)$ on the sensor $\mathrm{S}_b$, with $d$ the smallest length scale of the aperture function of the object. Thus, as far as the pixel sizes lie above the resolution limits, the spatial and angular resolution are decoupled. The structure of a standard plenoptic device, instead, entails an inverse proportionality relation between the angular resolution and the spatial resolution of the focused image, also in the geometrical-optics regime \cite{adelson,ng}. Thus, our protocol of plenoptic imaging with entangled photons enables us to beat this intrinsic limitation and achieve a larger depth of field (depending on the angular resolution), by leaving unchanged the resolution on the focused image and the total number of pixels.

\section{Simulation of CPI With Entangled Photons From SPDC}\label{sect:simulation}

\begin{figure}[H]
\centering
{
\includegraphics[width=0.31\textwidth]{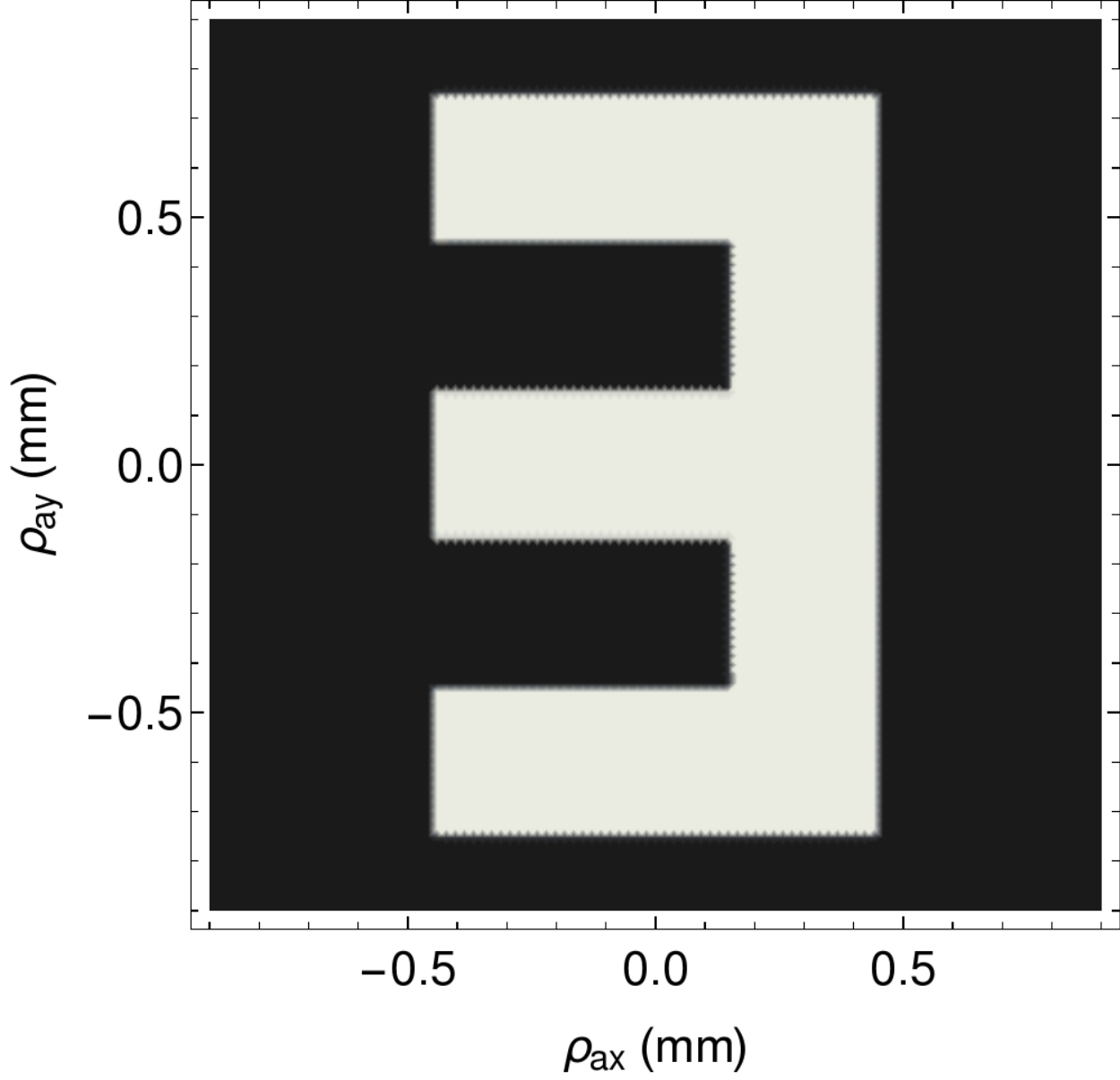}\quad
\includegraphics[width=0.30\textwidth]{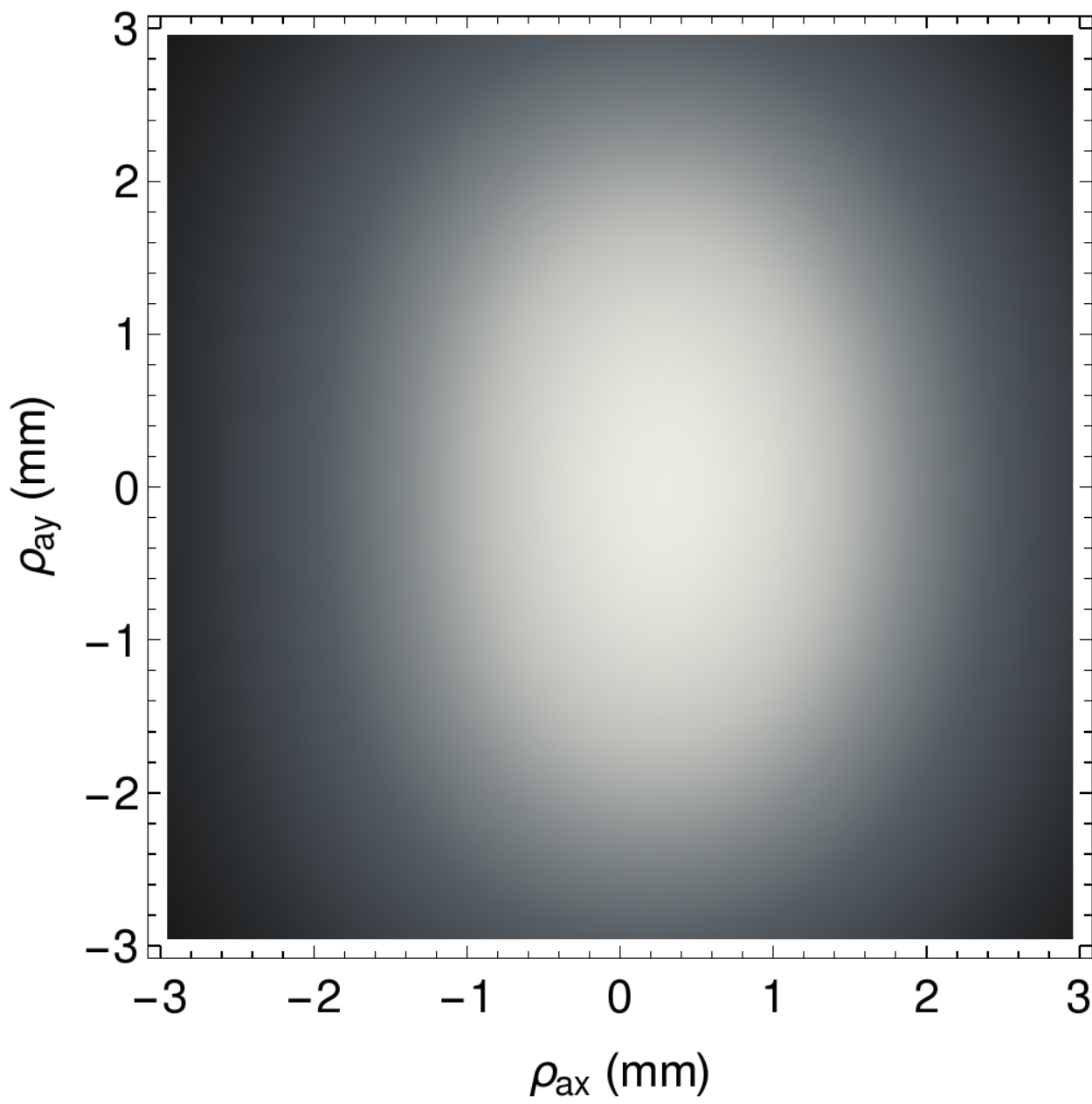}\quad
\includegraphics[width=0.31\textwidth]{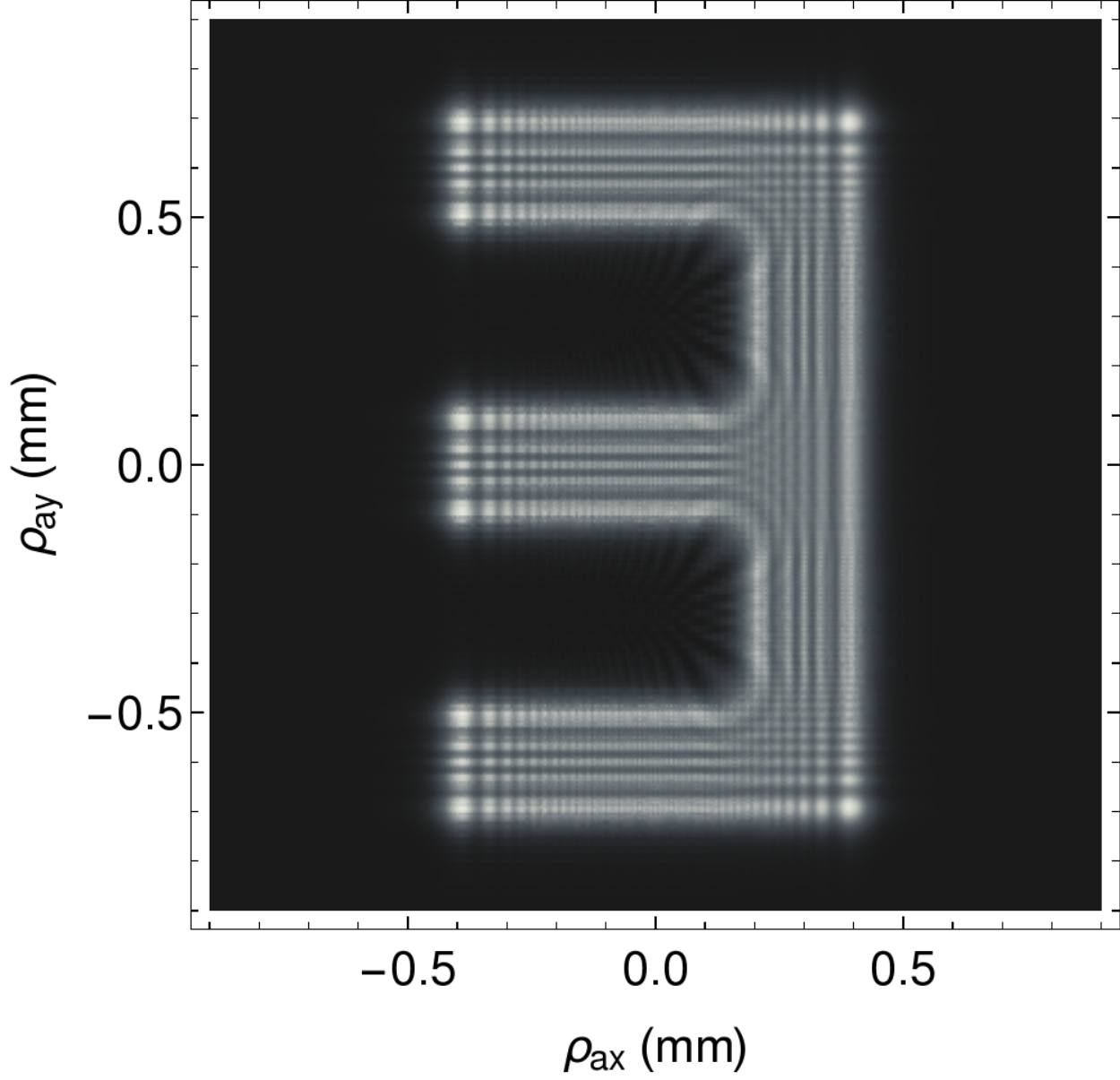}
\includegraphics[width=0.4\textwidth]{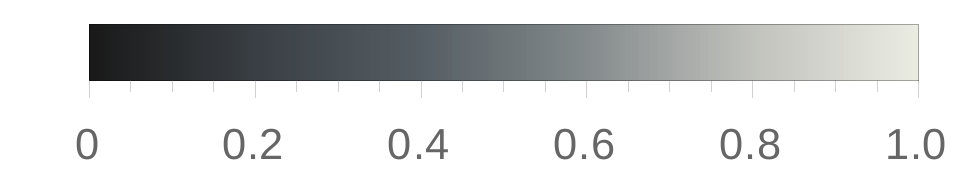}}
\caption{Comparison between focused (left), misfocused (center) and refocused (right) images of a two-dimensional object. Intensities are normalized to their maximum value in all panels. The out-of-focus and the refocused images are taken in the \textit{same} setup, in which $z_b \approx z_{bF}/3$.  }\label{fig:refocusing}
\end{figure}

In Figure \ref{fig:refocusing}, we show the enhanced depth of field induced by the refocusing capability of the SPDC correlation plenoptic protocol. A mask with a transparent letter $E$, whose thickness is\mbox{ $d=0.2\,\mathrm{mm}$,} is placed in a setup with $z_a=10\,\mathrm{mm}$, $z_a'=30\,\mathrm{mm}$, and $f=12\,\mathrm{mm}$, which would give a focused ghost image magnified by $m=1.5$. The object mask is illuminated by SPDC photons with $\lambda=1\, \mu\mathrm{m}$, generated by a pump whose Gaussian transverse profile has width $\sigma=0.6\,\mathrm{mm}$. With respect to the source, the object is placed at a distance $z_b=3\,\mathrm{mm}$, which is less than one third of the focused plane distance $z_{bF}=10\,\mathrm{mm}$. The ghost image of such an object would be focused at $z_{aF}'=5 z_a'$. The widths of the sensors $\mathrm{S}_a$ and $\mathrm{S}_b$ are fixed to $W_a=6 m d =1.8\,\mathrm{mm}$ and $W_b = 4 M \sigma = 1.9\,\mathrm{mm}$, with $M=0.8$ the magnification of the source image reproduced on $\mathrm{S}_b$. Their pixel size $\delta=6\,\mu\mathrm{m}$ is close to both resolution limits, as defined by the source and the object's aperture. The results reported in Figure \ref{fig:refocusing} clearly indicate that the refocusing procedure enables the recovery of the information on the aperture function of the object, which is completely lost in the misfocused \mbox{ghost image. }

We shall now compare the above results with the one achievable by a standard plenoptic camera having the same pixel size and total number of pixels per side ($N_{\mathrm{tot}}=N_a+N_b=620$). To this end, we introduce the parameter $\alpha=S_i/S'_i$, given by the ratio between the distance $S_i$ from the focusing element to the image plane, and the actual distance $S'_i$ between the focusing element (imaging lens) and the detector. Generally, perfect refocusing is possible if \cite{ng}
\begin{equation}\label{ref}
\left| 1-\frac{1}{\alpha} \right| < \frac{\Delta x}{\Delta u}, 
\end{equation}
where $\Delta x$ is the minimum distance that can be resolved on the image plane, and $\Delta u$ the minimum distance that can be resolved on the imaging lens. In a standard plenoptic camera, if the sensor have pixels of size $\delta$, the image resolution is given by $\Delta x^{(\mathrm{p})}= 2 \delta N_u^{(\mathrm{p})}$, while each pixel $\delta$ coincides with an area of width $\Delta u^{(\mathrm{p})} = 2 D_s/N_u^{(\mathrm{p})}$ on the lens, with $D_s$ the lens diameter. Hence,
\begin{equation}\label{DOF_p}
\left( \frac{\Delta x}{\Delta u} \right)^{(\mathrm{p})} = \frac{\delta}{D_s} \left( N_u^{(\mathrm{p})} \right)^2.
\end{equation}

In CPI instead, $\Delta x^{(\mathrm{c})}=2 \delta$, since pixels of width $\delta$ can be used also to retrieve the image. On the other hand, the resolution on the imaging lens is given by $\Delta u^{(\mathrm{c})}= 2 D_s/N_b$, where $D_s$ is the effective diameter of the lens $L_a$, that can be obtained by properly scaling the size $D'_s$ of the pump profile:
\begin{equation}
D_s = D'_s \left( 1+ \frac{z_a}{z_b} \right).
\end{equation} 

In this case, the right-hand side of the perfect refocusing condition given in Equation (\ref{ref}) reads
\begin{equation}\label{DOF_cp}
\left( \frac{\Delta x}{\Delta u} \right)^{(\mathrm{c})} =
\frac{\delta}{D_s} N_b.
\end{equation}

Hence, the maximum achievable depth of focus, in the setup employed for the simulation reported in Figure \ref{fig:refocusing}, is $|1-1/\alpha|<0.26$. A standard plenoptic camera with the same pixel size and total number of pixels per side would enable us to achieve this same depth of focus provided $N_u^{\mathrm{(p)}}=18$ pixels are employed for the angular resolution; this condition imposes a loss of spatial resolution by a factor $18$ ($\Delta x^{\mathrm{(p)}}=0.1\,\mathrm{mm}$) with respect to the one of the CPI protocol considered above.

%%%%%%%%%%%%%%%%%%%%%%%%%%%%%%%%%%%%%%%%%%
\section{Discussion}

\begin{figure}[H]
\centering
{\includegraphics[width=0.4\textwidth]{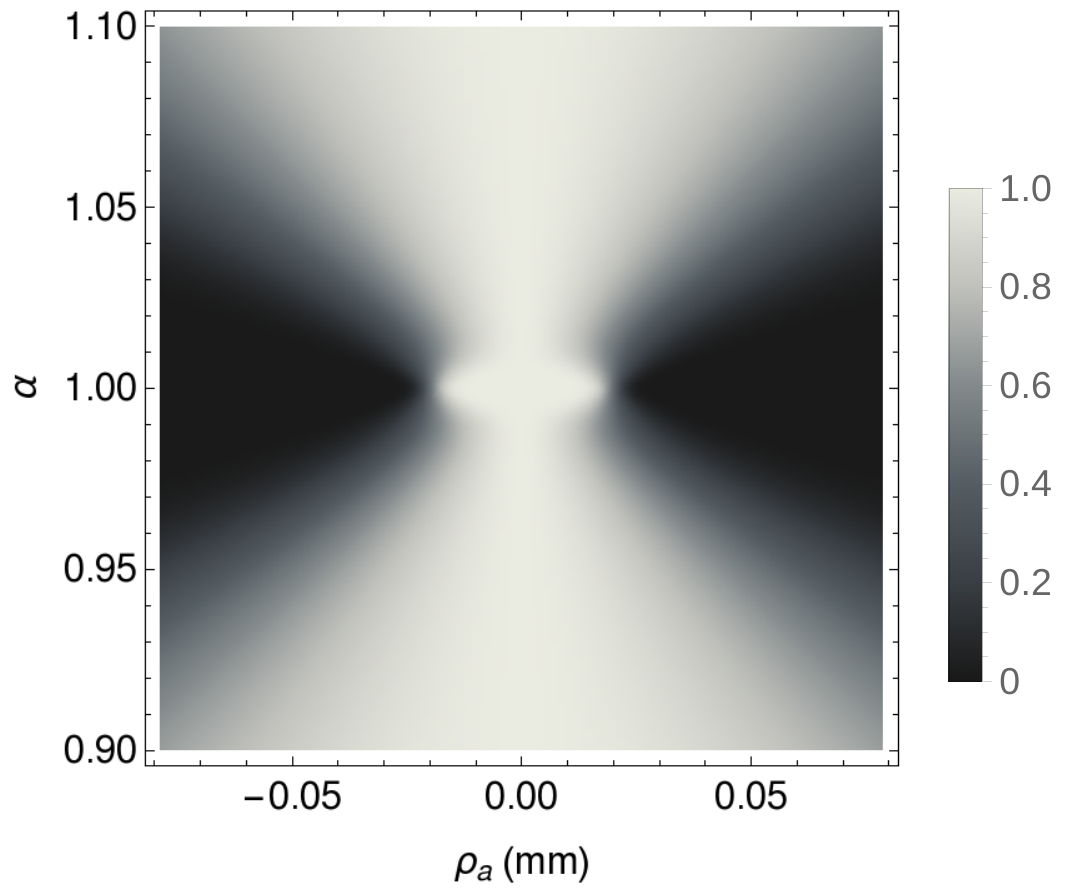}
\includegraphics[width=0.4\textwidth]{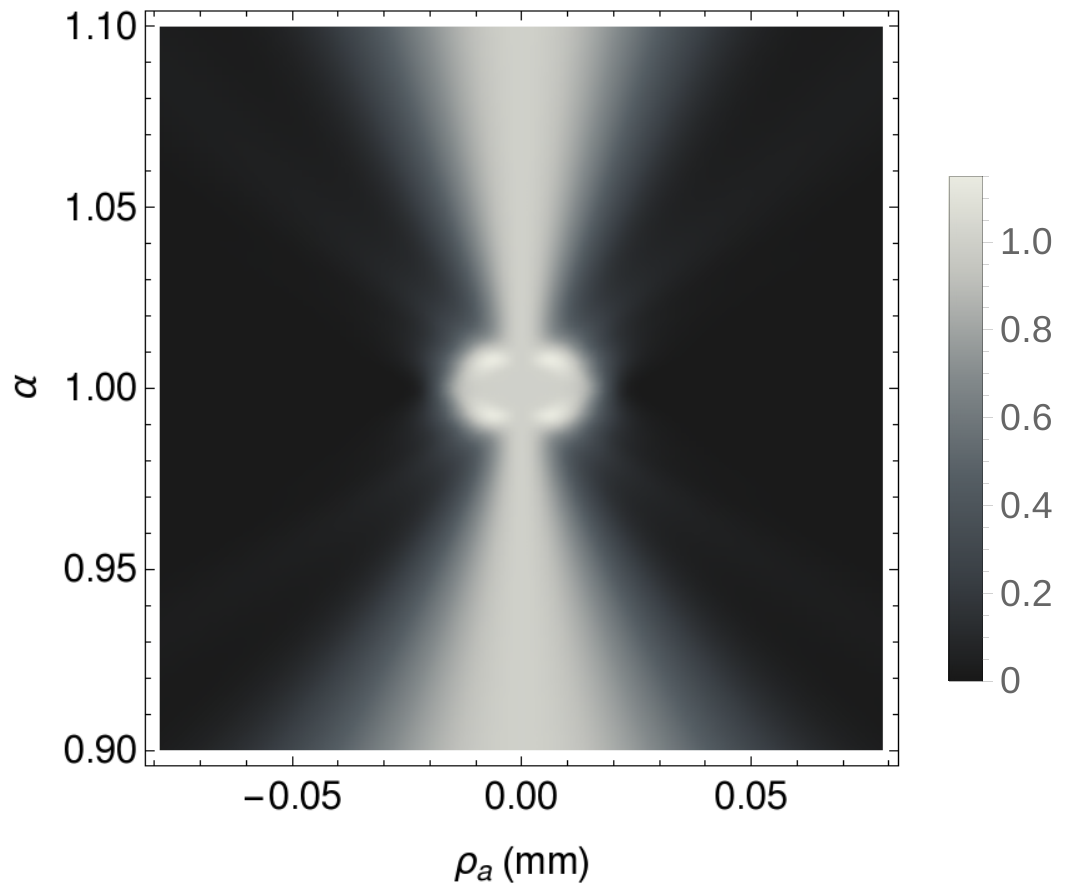}}
\caption{Comparison between the coherent and the incoherent ghost image of a single slit of width $a=26\,\mu\mathrm{m}$, as given by Equations~(\ref{Sigma_GI}) and (\ref{Gamma_cGI}), respectively, in the same setup described in Section~\ref{sect:simulation}. Both functions have been normalized to their value in $\rho_{a}=0$ for any value of $\alpha$. }\label{fig:coh_vs_incoh}
\end{figure}

At the heart of the refocusing capability of the second order correlation function of Equation (\ref{Gamma}), is the larger depth of focus of the coherent ghost image (Equation  (\ref{Gamma_cGI})), with respect to the incoherent ghost image (Equation (\ref{Sigma_GI})), as reported in Figure \ref{fig:coh_vs_incoh}.  In fact, the maximum achievable depth of focus of the proposed CPI scheme is the result of the increased depth of focus of coherent ghost imaging, with respect to incoherent ghost imaging.

\begin{figure}[H]
\centering
{\includegraphics[width=0.75\textwidth]{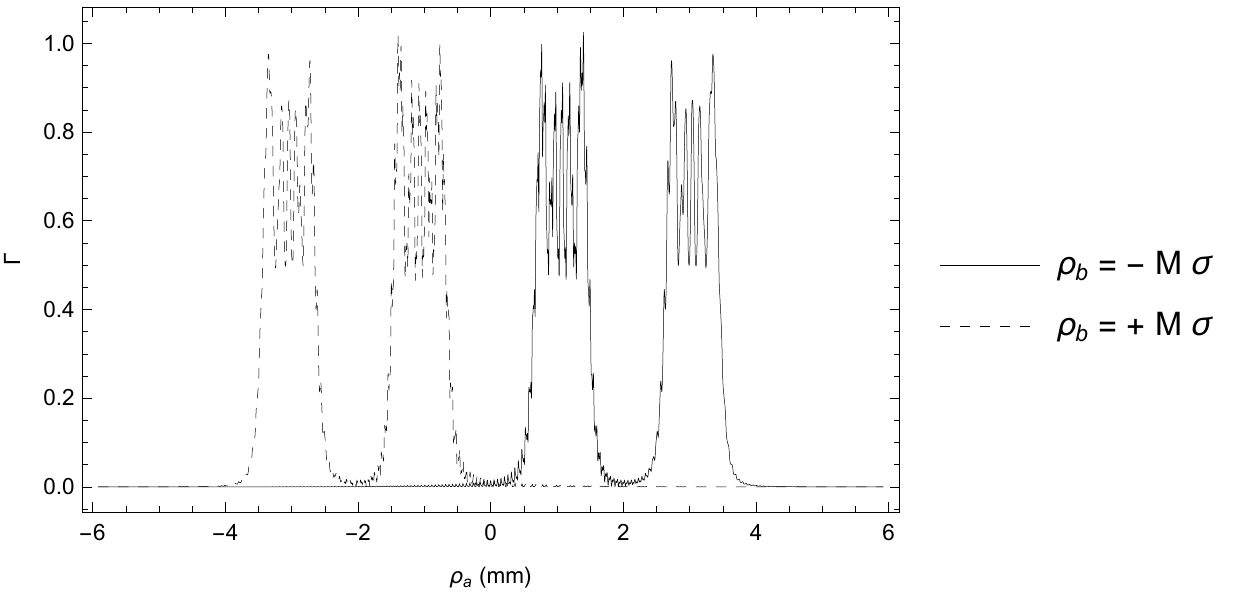}}
\caption{Observation of a double slit of width $a=0.2\,\mathrm{mm}$ and center-to-center distance $2a$ from two different points of view. Here, the setting is one-dimensional, with the same parameters as the setup described in Section \ref{sect:simulation}. The coherent ghost images of Equation~(\ref{Gamma}) enable us to change the point of view on any out-of-focus plane by selecting the point $\rho_b$ on the sensor $\mathrm{S}_b$, corresponding to a source point $\rho_s=-\rho_b/M$. In this case, the axis of the double slit coincides with the optical axis, and the chosen points on $\mathrm{S}_b$ are $\rho_b=- M\sigma$ (solid line, on the right) and $\rho_b=+M\sigma$ (dashed line, on the left).}\label{fig:cohx2}
\end{figure}

This can be better understood by considering the origin of both the out-of-focus and the refocused image: The first one is obtained by integrating the out-of-focus coherent image\mbox{ (Equations (\ref{Gamma}), (\ref{Gamma_phi}), or (\ref{Gamma_out})) }over the whole sensor $\mathrm{S}_b$, exactly as it would do a bucket detector of standard ghost imaging; the second one is obtained by integrating, over the same sensor $\mathrm{S}_b$, the rescaled version of such out-of-focus coherent image, as indicated in Equation (\ref{Sigma_riscaled}). Now, as shown in Figure \ref{fig:cohx2}, the out-of-focus coherent image is a projection of the focused image (hence, it is either enlarged or reduced with respect to it) as seen by the \textit{viewpoint} defined by the specific value of $\bm{\rho}_b$. The integration all such coherent images over the whole sensor $\mathrm{S}_b$ implies the overlap of all the projections taken from the different viewpoints $\bm{\rho}_b$; the resulting incoherent image is thus characterized by a loss of resolution, namely, it appears out of focus. The rescaled coherent image restores the correct size of the focused image and, most important, \textit{tilts} the image in such a way to cancel the specific viewpoint from which it was taken. As a consequence, the integration of all such rescaled coherent images over the whole sensor $\mathrm{S}_b$ has no more detrimental effect on the resolution of the resulting incoherent image; the post-processed image thus appears refocused.

\section{Conclusions and Outlook}

In view of practical applications, it is worth mentioning that all the above results apply to both reflective and transmitting objects. In addition, in contrast with chaotic light, entangled photons from SPDC enable us to employ different wavelengths in the two arms of the setup: Light illuminating the object is not required to have the same spectrum as light being remotely detected by $\mathrm{S}_a$ to retrieve the desired image \cite{GI-2colour_th,GI-2colour_exp}. This is quite interesting in view of applications requiring specific illumination wavelenghts for the object. In this scenario, one may choose two different sensors for maximizing the detection efficiency.

As plenoptic imaging is being broadly adopted in diverse fields such as digital \mbox{photography~ \cite{website1,website2,website3}}, microscopy \cite{microscopy2,microscopy4}, 3D imaging, sensing and rendering \cite{3dimaging}, our proposed scheme has direct applications in several biomedical and engineering fields. Interestingly, the coherent nature of the correlation plenoptic imaging technique may lead to innovative coherent microscopy modality.

%................................................................................................................................................
%% April 1st special: add yet another coffee stain
%\cofeBm{0.7}{1}{0}{0}{0}
%................................................................................................................................................

%%%%%%%%%%%%%%%%%%%%%%%%%%%%%%%%%%%%%%%%%%

%%%%%%%%%%%%%%%%%%%%%%%%%%%%%%%%%%%%%%%%%%
\vspace{6pt} 

%%%%%%%%%%%%%%%%%%%%%%%%%%%%%%%%%%%%%%%%%%
\acknowledgments{This work has been supported by the MIUR project P.O.N.~RICERCA E COMPETITIVITA' 2007-2013 - Avviso n.~713/Ric.~del 29/10/2010, Titolo II - ``Sviluppo/Potenziamento di DAT e di LPP'' (\mbox{project n.~PON02-00576-3333585}), the INFN through the project ``QUANTUM'', the UMD Tier 1 program and the Ministry of Science of Korea, under the ``ICT Consilience Creative Program'' (IITP-2015-R0346-15-1007).}

%%%%%%%%%%%%%%%%%%%%%%%%%%%%%%%%%%%%%%%%%%
\authorcontributions{Milena D'Angelo, Giuliano Scarcelli and Augusto Garuccio conceived and designed the proposed scheme for CPI; Francesco V. Pepe performed the theoretical calculation; Francesco Di Lena performed the simulation; Milena D'Angelo and Francesco V. Pepe wrote the paper. All authors have read and approved the final manuscript.}

%%%%%%%%%%%%%%%%%%%%%%%%%%%%%%%%%%%%%%%%%%
\conflictofinterests{The authors declare no conflict of interest. The founding sponsors had no role in the design of the study; in the collection, analyses, or interpretation of data; in the writing of the manuscript, and in the decision to publish the results.} 

%%%%%%%%%%%%%%%%%%%%%%%%%%%%%%%%%%%%%%%%%%
\bibliographystyle{mdpi}

%=====================================
% References, variant A: internal bibliography
%=====================================
\renewcommand\bibname{References}

%=====================================
% References, variant B: external bibliography
%=====================================
%\bibliography{your_external_BibTeX_file}

%%%%%%%%%%%%%%%%%%%%%%%%%%%%%%%%%%%%%%%%%%
\end{document}